\RequirePackage{arydshln}
\documentclass[aps,twocolumn,nofootinbib,superscriptaddress,preprintnumbers,pra,10pt]{revtex4-1}

\usepackage{float}
\usepackage{colortbl}
\usepackage{amsmath,amssymb}
\usepackage{dsfont} 
\usepackage{hyperref}
\usepackage{graphicx}
\usepackage{enumitem}
\usepackage{arydshln}
\usepackage{mathtools}
\usepackage{bbold}
\usepackage{multirow}

\makeatletter
\g@addto@macro\bfseries{\boldmath}
\makeatother

\newcommand{\be}{\begin{equation}}
\newcommand{\ee}{\end{equation}}
\newcommand{\bea}{\begin{eqnarray}}
\newcommand{\eea}{\end{eqnarray}}

\newcommand{\cO}{{\mathcal O}}
\newcommand{\cC}{{\mathcal C}}
\newcommand{\cL}{{\mathcal L}}

\newcommand{\cA}{{\mathcal A}}
\newcommand{\cB}{{\mathcal B}}
\newcommand{\Tr}{{\rm Tr}}
\newcommand{\Vq}{V_q}
\newcommand{\Vl}{V_\ell}
\newcommand{\Vql}{V_{q,\ell}}
\newcommand{\DD}{\Delta}
\newcommand{\gsim}{\lower.7ex\hbox{$\;\stackrel{\textstyle>}{\sim}\;$}}
\newcommand{\lsim}{\lower.7ex\hbox{$\;\stackrel{\textstyle<}{\sim}\;$}}

\makeatletter
\renewcommand\paragraph{\@startsection{paragraph}{4}{\z@}%
                                    {3.25ex \@plus1ex \@minus.2ex}%
                                    {-1em}%
                                    {\normalfont\normalsize\bfseries}}
\makeatother

\allowdisplaybreaks

\begin{document}

\preprint{ZU-TH-42/19}

\title{With or without $U(2)$? Probing non-standard flavor \\ and helicity structures in 
semileptonic $B$ decays}

\author{Javier Fuentes-Mart\'{\i}n}
\email{fuentes@physik.uzh.ch}
\affiliation{Physik-Institut, Universit\"at Zu\"rich, CH-8057 Z\"urich, Switzerland}
\author{Gino Isidori}
\email{gino.isidori@physik.uzh.ch}
\affiliation{Physik-Institut, Universit\"at Zu\"rich, CH-8057 Z\"urich, Switzerland}
\author{Julie Pag\`es}
\email{juliep@physik.uzh.ch}
\affiliation{Physik-Institut, Universit\"at Zu\"rich, CH-8057 Z\"urich, Switzerland}
\author{Kei Yamamoto}
\email{keiy@hiroshima-u.ac.jp}
\affiliation{Physik-Institut, Universit\"at Zu\"rich, CH-8057 Z\"urich, Switzerland}
\affiliation{Graduate School of Science, Hiroshima University, Higashi-Hiroshima 739-8526, Japan}

\begin{abstract}
\vspace{5mm}
Motivated by the recent hints of  lepton flavor universality violation observed in semileptonic $B$ decays,
we analyze how to test flavor and helicity structures of the corresponding  amplitudes in view of future data.  
We show that the general assumption that such non-standard effects are controlled by a 
$U(2)_q \times U(2)_\ell$ flavor symmetry, minimally broken as in the Standard Model Yukawa sector, 
leads to stringent predictions on leptonic and semileptonic $B$ decays.  
Future measurements of  $R_{D^{(*)}}$, $R_{K^{(*)}}$, $\cB(\bar B_{c,u}\to \ell \bar \nu)$, $\cB(\bar B \to \pi \ell \bar\nu)$, $\cB(B \to \pi \ell \bar\ell)$, 
$\cB(B_{s,d}\to\ell\bar\ell^{(\prime)})$, 
as well as various polarization asymmetries in $\bar B\to D^{(*)} \tau \bar \nu$ decays, will allow to prove or falsify this general
hypothesis independently of its dynamical origin. 
\vspace{3mm}
\end{abstract}

\maketitle

\allowdisplaybreaks

\section{Introduction}\label{sec:intro}
Present data exhibit intriguing hints of violations of  Lepton Flavor Universality (LFU) 
both in charged-current~\cite{Aaij:2015yra,Lees:2013uzd,Hirose:2016wfn,Aaij:2017deq,Abdesselam:2019dgh} 
and neutral-current~\cite{Aaij:2014ora,Aaij:2017vbb,Aaij:2019wad,Wei:2009zv,Abdesselam:2019wac,Aubert:2006vb} 
semileptonic $B$ decays. These hints can be well described employing Effective Field Theory (EFT) approaches 
(see~\cite{Bhattacharya:2014wla, Alonso:2015sja, Greljo:2015mma, Calibbi:2015kma, Barbieri:2015yvd} for the early attempts),
whose main ingredients are the assumptions that New Physics (NP) affects predominantly semileptonic operators, 
and that it couples in a non-universal way to  different fermion species.
In particular, NP should have dominant couplings to   
third generation fermions and smaller, but non-negligible, couplings to  second generation fermions.
Interestingly enough, this non-trivial flavor structure resemble the hierarchies observed in the 
Standard Model (SM) Yukawa couplings, opening the possibility of a common explanation for the two phenomena. 

An effective approach to address the question of a possible connection between  
the LFU anomalies and the SM Yukawa couplings and, more generally, to investigate 
the flavor structure of  non-standard effects at low-energies, is the assumption 
of an appropriate flavor symmetry and a set of symmetry-breaking terms. 
The flavor symmetry does not need to be a fundamental property of the underlying theory,
it could simply be an accidental low-energy property. Still, its use in the EFT context provides a very
powerful organizing principle for a bottom-up reconstruction of the underlying dynamics. 

In the context of the recent anomalies, the flavor symmetry that emerges as particularly 
suitable to describe the observed data is $U(2)_q \times U(2)_\ell$, which is a subset 
of the larger $U(2)^5$ proposed in~\cite{Barbieri:2011ci,Blankenburg:2012nx, Barbieri:2012uh} as a useful 
organizing principle to address the hierarchies in the SM Yukawa couplings and, at the same time,
allow large NP effects in processes involving third-generation fermions (as expected 
in most attempts to address the electroweak hierarchy problem).  

The scope of the present paper is a systematic investigation of the consequences of this symmetry hypothesis
in (semi)leptonic $B$ decays. Contrary to previous analyses,
where this symmetry has been implemented in the context of specific new-physics models, 
our goal here is to investigate the consequences of this symmetry (and symmetry-breaking) ansatz 
in general terms,  with  a minimal set of additional assumptions about the dynamical origin of the anomalies.
As we will show, in the case of charged-current interactions, the symmetry ansatz alone is sufficient to 
derive an interesting series of testable predictions. The predictive power is smaller for neutral-current transitions,
but also in that case we can identify a few clean predictions which are direct consequences of the 
symmetry ansatz alone.

\section{The $U(2)^5$ symmetry in the SM}\label{sec:U2}
The  $U(2)^5\times U(1)_{B_3}\times U(1)_{L_3}$ symmetry is the global symmetry that the SM Lagrangian exhibits in the limit where 
we neglect all entries in the Yukawa couplings but for third generation masses~\cite{Barbieri:2011ci,Blankenburg:2012nx, Barbieri:2012uh}.
Under this symmetry, the first two SM fermion families transform as doublets of a given $U(2)$ subgroup,
\begin{align}
U(2)^5\equiv U(2)_{q}\times  U(2)_\ell \times U(2)_u \times U(2)_d \times U(2)_e\,,
\end{align}
while third-generation quarks (leptons) are only charged under $U(1)_{B_3(L_3)}$. The largest breaking of this symmetry in the 
complete SM Lagrangian is controlled by the small parameter 
\begin{align}
\epsilon =   \left[    \Tr(Y_u Y_u^\dagger) - \frac{ \Tr(Y_u Y_u^\dagger Y_d Y_d^\dagger)}{\Tr(Y_d Y_d^\dagger)}    \right]^{1/2} 
\approx y_t |V_{ts}| \approx 0.04\,.
\end{align}

A minimal set of $U(2)^5$ breaking terms (\textit{spurions}) 
which lets us reproduce all the observable SM flavor parameters
(in the limit of vanishing neutrino masses), 
without tuning and with minimal size for the breaking terms, is  
\begin{align}
\begin{aligned}
\Vq&\sim\left(\boldsymbol{2},\boldsymbol{1},\boldsymbol{1},\boldsymbol{1},\boldsymbol{1}\right)\,,& \Vl&\sim\left(\boldsymbol{1},\boldsymbol{2},\boldsymbol{1},\boldsymbol{1},\boldsymbol{1}\right)\,,\\ 
\DD_{u(d)}&\sim\left(\boldsymbol{2},\boldsymbol{1},\boldsymbol{\bar 2}(\boldsymbol{1}),\boldsymbol{1}(\boldsymbol{\bar 2}),\boldsymbol{1}\right)\,,&\DD_e&\sim\left(\boldsymbol{1},\boldsymbol{2},\boldsymbol{1},\boldsymbol{1},\boldsymbol{\bar 2}\right) \,.
\end{aligned}
\end{align}
In terms of these spurions, the $3\times 3$ Yukawa matrices can be decomposed as
\begin{align}
\label{eq:U2Yukawas}
\begin{aligned}
Y_{u (d)}&=y_{t(b)}
\begin{pmatrix}
\DD_{u(d)} & x_{t(b)}\,\Vq\\
0 & 1
\end{pmatrix}
\,,&
Y_e&=y_\tau
\begin{pmatrix}
\DD_e & x_\tau\,\Vl\\
0 & 1
\end{pmatrix}
\,,
\end{aligned}
\end{align}
where $x_{t,b,\tau}$ and $y_{t,b,\tau}$ are free complex parameters, expected to be of $\cO(1)$.\footnote{In 
models with more than one Higgs doublet, the smallness of $y_{b,\tau}$ can be justified in terms of approximate 
flavor-independent $U(1)$ symmetries.} Note that $\Delta_{u,d,e}$ 
are $2\times2$ complex matrices, while $V_{q,\ell}$ are 2-dimensional complex vectors.

The precise size of the spurions is not known; however, we can estimate it by the requirement of no tuning in the 
$\cO(1)$ parameters. This implies $|\Vq| =O(\epsilon)$. In the limit of vanishing neutrino masses, the size of $|\Vl|$  cannot be unambiguously determined.
As discussed below (see also~\cite{Cornella:2019hct}), a good fit of the anomalies in semileptonic $B$ decays is obtained for 
\begin{align}\label{eq:leadingV}
|\Vl|,~|\Vq| = O(10^{-1})~,
\end{align}
which is perfectly consistent with: i) the estimate $|\Vq| =O(\epsilon)$;
ii) the hypothesis of a common origin for the two leading $U(2)^5$ breaking terms in quark and lepton sectors.
The entries in the $2\times 2$ matrices $\DD_{u,d,e}$ are significantly smaller than $|V_{q,\ell}|$, with a maximal 
size of $O(10^{-2})$  in the quark sector.

By appropriate field redefinitions and without loss of generality, one can remove unphysical parameters in the Yukawa matrices in~\eqref{eq:U2Yukawas} (see App.~\ref{app:U2Yukawas}).
 Working in the so-called interaction basis, where the second generation in $U(2)_{q(\ell)}$ space is defined by the alignment of the leading spurions,
\begin{align}
 \Vql &=  |\Vql| \times \vec n\,, \qquad  \vec n=\begin{pmatrix}0 \\ 1\end{pmatrix}~,
\end{align}
one can bring the Yukawa matrices to the following form
\begin{align}
\label{eq:U2YukawasNonR}
\begin{aligned}
Y_u&=|y_t|
\begin{pmatrix}
U_q^\dagger O_u^\intercal\, \hat\DD_u && |V_q|\,|x_t|\,e^{i\phi_q}\,\vec n\\
0 && 1
\end{pmatrix}
\,,\\
Y_d&=|y_b|
\begin{pmatrix}
\;\;\;U_q^\dagger \hat\DD_d&&& |V_q|\,|x_b|\,e^{i\phi_q}\,\vec n\\
\;\;\;0 &&& 1
\end{pmatrix}
\,,\\
Y_e&=|y_\tau|
\begin{pmatrix}
\;\;O_e^\intercal\,\hat\DD_e\;\;&&&& |V_\ell|\,|x_\tau|\,\vec n\\
\;\;0 &&&& 1
\end{pmatrix}
\,,
\end{aligned}
\end{align}
where $\hat\DD_{u,d,e}$ are $2\times 2$ diagonal positive matrices, $O_{u,e}$ are $2\times2$ orthogonal matrices and $U_q$ is of the form
\begin{align}\label{eq:Uq}
U_q=
\begin{pmatrix}
c_d & s_d\,e^{i\alpha_d}\\
-s_d\,e^{-i\alpha_d} & c_d
\end{pmatrix}\,,
\end{align}
with $s_d\equiv\sin\theta_d$ and $c_d\equiv\cos\theta_d$.

The Yukawa matrices in~\eqref{eq:U2YukawasNonR} get diagonalized by means of appropriate unitary transformations: 
$L_f^\dagger Y_f R_f  = {\rm diag}(Y_f)$, with $f=u,d,e$. The most general form for these unitary transformations is
\begin{align}\label{eq:YukRot}
\begin{aligned}
 L_d  &\approx 
\begin{pmatrix}
 c_d   &  -s_d\,e^{i\alpha_d}  & 0  \\
 s_d\,e^{-i\alpha_d}   &  c_d &  s_b   \\
-s_d\,s_b\,e^{-i(\alpha_d+\phi_q)}   & -c_d\,s_b\, e^{-i\phi_q} & e^{-i\phi_q}
\end{pmatrix}
\,,\\[5pt]
 R_d  &\approx  
\begin{pmatrix}
1   &  0  & 0  \\
0   &  1  & \frac{m_s}{m_b}\,s_b   \\
0  & -\frac{m_s}{m_b}\,s_b\,e^{-i\phi_q} & e^{-i\phi_q}
\end{pmatrix}
\,,\\[5pt]
 R_u  &\approx
\begin{pmatrix}
1   &  0  & 0  \\
0   &  1  & \frac{m_c}{m_t}\,s_t   \\
0  & -\frac{m_c}{m_t}\,s_t\,e^{-i\phi_q} & e^{-i\phi_q}
\end{pmatrix}
\,,\\[5pt]
 L_e  &\approx    
\begin{pmatrix}
c_e       & -s_e          & 0  \\
s_e &  c_e           & s_\tau \\
-s_e s_\tau   &  -c_es_\tau & 1
\end{pmatrix}\,,\\[5pt]
 R_e  &\approx  
\begin{pmatrix}
 1   & 0 & 0  \\
0 &  1& \frac{m_\mu}{m_\tau}\, s_\tau \\
0   &  -\frac{m_\mu}{m_\tau}\, s_\tau & 1
\end{pmatrix}\,,
\end{aligned}
\end{align}
with $L_u=L_d\,V_{\rm CKM}^\dagger$.  Here we have taken advantage of the constraints imposed by 
fermions masses and CKM matrix elements to eliminate various parameters appearing in $L_f$ and 
$R_f$.\footnote{The removal of  unphysical parameters presented in App.~\ref{app:U2Yukawas}
corrects a similar analysis presented in~\cite{Bordone:2018nbg},
where it was erroneously concluded that the parameter $\alpha_d$ is unconstrained.}
These further imply that $s_d$ and $\alpha_d$ are constrained by
$s_d/c_d=|V_{td}/V_{ts}|$ and $\alpha_d=\arg(V_{td}^*/V_{ts}^*)$, and that $s_t=s_b-V_{cb}$. 
The light-family leptonic mixing ($s_e$), appearing in $O_e$,  cannot be expressed  in terms of measurable quantities.
Two additional mixing angles which remain unconstrained are $s_b/c_b=|x_b|\,|V_q|$ and $s_\tau/c_\tau=|x_\tau|\,|V_\ell|$.
Finally, $\phi_q$ is an unconstrained $O(1)$ phase, that becomes unphysical in the limit $s_b\to 0$ (or, equivalently, $x_b\to0$). 
This limit is phenomenologically required in models where $\Delta F=2$ operators are generated at tree-level around the TeV scale: in such case 
one needs to impose a mild alignment to the down basis, i.e.~$|s_b| \lesssim 0.1\, \epsilon$, to satisfy the constraints from $B_{s,d}$-meson mixing~\cite{DiLuzio:2017vat,Bordone:2018nbg,DiLuzio:2018zxy}.

\section{Impact of $U(2)^5$ on the EFT for semileptonic $B$ decays}\label{sec:EFTU2}

Having defined the flavor symmetry and its symmetry breaking terms from the SM Yukawa sector,
we are ready to analyze its implications beyond the SM. 
Assuming no new degrees of freedom below the electroweak scale, we can describe NP effects in full generality 
employing the so-called  SMEFT. We limit the attention to dimension-six four-fermion operators bilinear in 
quark and lepton fields,\footnote{We neglect operators which modify the effective 
couplings of $W$ and $Z$ bosons. These are highly constrained and cannot induce sizable LFU violating effects.}
that we write generically as 
\begin{align}
\begin{aligned}
\cL_{\rm EFT}&=-\frac{1}{v^2}  \sum_{k,[ij\alpha\beta]} \cC^{[ij\alpha\beta]}_k  \cO^{[ij\alpha\beta]}_k +{\rm h.c.} \,,
\end{aligned}
\end{align}
where $v\approx 246$~GeV is the SM Higgs vev, $\{\alpha, \beta\}$ are lepton-flavor indices, and $\{ i, j \}$ are quark-flavor indices.
The operators in the Warsaw basis~\cite{Grzadkowski:2010es} with a non-vanishing 
tree-level matrix element in semileptonic $B$ decays are 
\begin{align}
\begin{aligned}
\cO_{\ell q}^{(1)}&=(\bar \ell^\alpha_L\gamma^\mu\ell^\beta_L)(\bar q^i_L\gamma_\mu q^j_L)\,,\\
\cO_{\ell q}^{(3)}&=(\bar \ell^\alpha_L\gamma^\mu\tau^I\ell^\beta_L)(\bar q^i_L\gamma_\mu\tau^I q^j_L)\,,\\
\cO_{\ell d}&=(\bar \ell^\alpha_L\gamma^\mu\ell^\beta_L)(\bar d^i_R\gamma_\mu d^j_R)\,,\\
\cO_{qe}&=(\bar q^i_L\gamma^\mu q^j_L)(\bar e^\alpha_R\gamma_\mu e^\beta_R)\,,\\
\cO_{ed}&=(\bar e^\alpha_R\gamma^\mu e^\beta_R)(\bar d^i_R\gamma_\mu d^j_R)\,,\\
\cO_{\ell e d q}   &=(\bar \ell^\alpha_L e^\beta_R)(\bar d^i_R q^j_L)\,,\\
\cO_{\ell e q u}^{(1)}&=   (\bar \ell_L^{a,\alpha} e^\beta_R)\epsilon_{ab}(\bar q_L^{a,i}  u^j_R)  \,,\\
\cO_{\ell e q u}^{(3)}&= (\bar \ell_L^{a,\alpha} \sigma_{\mu\nu}e^\beta_R)\epsilon_{ab}(\bar q_L^{b,i}  \sigma^{\mu\nu} u^j_R) \,,
\end{aligned}
\end{align}
where $\tau^I$ are the Pauli matrices and $\{a,b\}$ are $SU(2)_L$ indices. 
Our main hypothesis is to reduce the number of $\cC^{[ij\alpha\beta]}_k$ retaining only those corresponding to 
$U(2)^5$ invariant operators, up to the insertion of  one or two powers of the leading $U(2)_q \times U(2)_\ell$ spurions
in~\eqref{eq:leadingV}.

A first strong simplification arises by neglecting subleading spurions 
with non-trivial transformation properties under $U(2)_{u,d,e}$. Since we are interested in processes of the type $b\to c(u) \ell\bar\nu$ and  $b\to s(d)\ell\bar\ell^{(\prime)}$, this implies that only the operators $\cO_{\ell q}^{(1)}$, $\cO_{\ell q}^{(3)}$, $\cO_{qe}$ and $\cO_{\ell e d q}$ can yield a relevant contribution. Among those, $\cO_{qe}$ can significantly contribute at tree-level only to $b\to s\tau\bar\tau$ transitions: since the latter are currently poorly constrained (see sect.~\ref{sec:neutral}), we do not consider this operator for simplicity.
We are thus left with the following effective Lagrangian
\begin{align}
\begin{aligned}
\label{eq:SMEFTLag}
\cL_{\rm EFT} &=-\frac{1}{v^2}\left[C_{V_1}\,\Lambda_{V_1}^{[ij\alpha\beta]}\, \cO_{\ell q}^{(1)} + C_{V_3}\,\Lambda_{V_3}^{[ij\alpha\beta]}\, \cO_{\ell q}^{(3)}\right.\\
&\qquad\left.+(2\,C_S\,\Lambda_S^{[ij\alpha\beta]}\,\cO_{\ell e d q}+{\rm h.c.})\right] \,,
\end{aligned}
\end{align}
where $C_{V_i,S}$  control the overall strength of the NP effects and  $\Lambda_{V_i,S}$ are tensors that parametrize the flavor structure.
They are normalized by setting $\Lambda_{V_i,S}^{[3333]}=1$, which is the only term surviving in the exact $U(2)^5$ limit.

Let us consider first the structure of $\Lambda^{[ij\alpha\beta]}_S$, which is particularly simple. Neglecting $U(2)_{d,e}$ breaking spurions,
it factorizes to 
\be
\Lambda^{[ij\alpha\beta]}_S = (\Gamma_L^\dagger)^{\alpha j}   \times  \Gamma_R^{i \beta}~,
\label{eq:LambdaSfact}
\ee
where, in the interaction basis, 
\be
\Gamma_L^{i\alpha} = 
\begin{pmatrix}
 x_{q\ell}  \Vq^i (\Vl^\alpha)^*  &   x_q  \Vq^i  \\[4pt]
 x_\ell (\Vl^\alpha)^*  & 1
\end{pmatrix}~, \quad 
 \Gamma_R = 
\begin{pmatrix}
0 & 0 \\
0 & 1
\end{pmatrix}~.
\label{eq:beta0}
\ee
Here $ x_{q,\ell,q\ell}$ are $O(1)$ coefficients and we have neglected higher-order 
terms in $\Vql$ (that would simply redefine such coefficients).
Moving to the mass-eigenstate basis of down quarks and charged leptons, 
where 
\begin{align}\label{eq:DownBasis}
q_L^i=
\begin{pmatrix}
V_{ji}^*\,u^j_L\\
d_L^i
\end{pmatrix}
\,,\qquad\qquad 
\ell_L^\alpha=
\begin{pmatrix}
\nu^\alpha_L\\
e_L^\alpha
\end{pmatrix}~,
\end{align}
we have $\Gamma_L  \to \hat \Gamma_L\equiv L_d^\dagger\, \Gamma_L L_e$ and $ \Gamma_R\to\hat\Gamma_R\equiv R_d^\dagger\, \Gamma_R R_e$
[see \eqref{eq:YukRot}],  with the new matrices assuming
the following explicit form in $3\times 3$ notation
\begin{align}\label{eq:betad}
\hat\Gamma_L & = e^{i\phi_q}
\begin{pmatrix}
\DD_{q\ell}^{d e}  & \DD_{q\ell}^{d \mu} &   \lambda_q^{d}  \\[3pt]
\DD_{q\ell}^{s e}   & \DD_{q\ell}^{s \mu}   & \lambda_q^{s}\\[4pt]
\lambda_\ell^{e}  &  \lambda_\ell^{\mu}  & x_{q\ell}^{b\tau}
\end{pmatrix}
\approx e^{i\phi_q}
\begin{pmatrix}
0  & 0  &   \lambda_q^{d}  \\[3pt]
0 & \DD_{q\ell}^{s \mu}    & \lambda_q^{s}\\[4pt]
\lambda_\ell^{e}  &  \lambda_\ell^{\mu}  & 1
\end{pmatrix}\,, \nonumber \\
\hat\Gamma_R  &\approx  e^{i\phi_q}
\begin{pmatrix}
0  & 0  &  0  \\[3pt]
0 &  0    & -\frac{m_s}{m_b}\,s_b\\[4pt]
0  &  -\frac{m_\mu}{m_\tau}\,s_\tau  & 1
\end{pmatrix}\,.
\end{align}
The (complex) parameters $x_{q\ell}^{b\tau}$, $\lambda_q^{i}$,  $\lambda_\ell^{\alpha}$, and  $\DD_{q\ell}^{\alpha i}$  are a combination of the 
spurions in (\ref{eq:beta0}) and the rotation terms from $L_{d,e}$, that satisfy
\begin{align}\label{eq:lambda_ratios}
 \lambda_q^{s} &= O(|\Vq|)\,,  & 
 \lambda_\ell^{\mu} &= O(|\Vl|)\,,    \nonumber\\
 x_{q\ell}^{b\tau} & = O(1)\,,  &  
 \DD_{q\ell}^{s \mu}  &= O( \lambda_q^{s} \lambda_\ell^{\mu} ) \,,  \nonumber\\
 \frac{\lambda_q^{d}}{ \lambda_q^{s} }&= \frac{\DD_{q\ell}^{d \alpha}}{ \DD_{q\ell}^{s \alpha} } =  \frac{V_{td}^*}{V_{ts}^*}\,, & 
\frac{\lambda_\ell^{e}}{ \lambda_\ell^{\mu}}  &=   \frac{\DD_{q\ell}^{i e}}{ \DD_{q\ell}^{i \mu} } = s_e \,.  
\end{align}
On the r.h.s.~of the first line of~(\ref{eq:betad}) we have neglected tiny terms suppressed by 
more than two powers of $|\Vql|$  or $s_{d,e}$.

If we consider at most one power of $\Vq$ and one power of  $\Vl$, then 
also $\Lambda^{[ij\alpha\beta]}_{V_i}$ factorizes into 
\be
\Lambda^{[ij\alpha\beta]}_{V_i} = ({\Gamma_L^{V_i}}^\dagger)^{\alpha j}   \times (\Gamma_L^{V_i})^{i \beta}\,,
\label{eq:LambdaVfact}
\ee
where $\Gamma_L^{V_1}$ and $\Gamma_L^{V_3}$ have  the same structure as $\Gamma_L$ with, a priori, 
different $O(1)$ coefficients for the spurions. Moving to the basis~\eqref{eq:DownBasis}, $\hat\Gamma_L^{V_i}$
assumes the same structure as $\hat\Gamma_L$ in~\eqref{eq:betad}, 
with parameters which can differ by O(1) overall factors,
but that obey the same flavor ratios as in~\eqref{eq:lambda_ratios}.  Corrections to the factorized structure in~\eqref{eq:LambdaVfact}
arises only to second order in $V_q$ or $V_\ell$, generating terms which are either 
irrelevant or can be reabsorbed in a redefinition of the observable parameters in the processes 
we are interested in (see sect.~\ref{sect:Obs}).

\subsection{Matching to the $U_1$ leptoquark case}

The EFT in~\eqref{eq:SMEFTLag}, with factorized flavor couplings as in~\eqref{eq:LambdaSfact} and \eqref{eq:LambdaVfact}, nicely matches the 
structure generated by integrating out a $U_1$ vector leptoquark, 
transforming as $ (\mathbf{3},\mathbf{1})_{2/3}$ under the SM gauge group.  
As noted first in \cite{Barbieri:2015yvd}, this field  provides indeed an 
excellent mediator to build in a natural, and sufficiently general way, an EFT 
for semileptonic $B$ decays built on the $U(2)^5$ flavor symmetry  
broken only by the leading $\Vq$  and $\Vl$ spurions (see \cite{Bhattacharya:2016mcc,DiLuzio:2017vat,Hiller:2017bzc,Buttazzo:2017ixm,Bordone:2018nbg,Angelescu:2018tyl,Kumar:2018kmr,DiLuzio:2018zxy,Aebischer:2019mlg,Cornella:2019hct} for other phenomenological analysis of the $U_1$ leptoquark in $B$ physics).

Writing the interaction between the $U_1$  field and SM fermions in the basis of~\eqref{eq:DownBasis} as~\cite{Cornella:2019hct}
\begin{align}
\mathcal L_{U_1}
=\frac{g_U}{\sqrt{2}}  \left[ \beta_{L}^{i\alpha}\,(\bar q_{L}^{\,i} \gamma_{\mu}  \ell_{L}^{\alpha})  +     \beta_{R}^{i \alpha }\,(\bar d_{R}^{\,i}\gamma_{\mu}   e_{R}^{\alpha})\right] U_1^\mu
+ {\rm h.c.}  \,,
\end{align}
the flavor symmetry hypothesis imply  a parametric structure for $\beta_L^{j\alpha}$ and  $\beta_R^{i\alpha}$
identical to that of  $\hat \Gamma_L^{i\alpha}$ and $\hat \Gamma_R^{i\alpha}$ in~\eqref{eq:betad}.  Normalizing $g_U$ such that $\beta_L^{b \tau} = 1$,
and integrating out the $U_1$ field, leads to the following (tree-level) 
matching conditions for the parameters of  $\cL_{\rm EFT}$
\begin{align}
\Gamma_L^{V_1} &= \Gamma_L^{V_3} = \Gamma_L \,, &  C_V&\equiv C_{V_1} = C_{V_3}= \frac{g_U^2 v^2}{4 M_U^2}>0 \,,
\label{eq:U1cond1}
\end{align}
and 
\begin{align}
\begin{aligned}
\frac{C_S}{C_V} &= - 2\, \beta_R~, &  \lambda_q^{s} &= \beta_L^{s \tau}~, &   \lambda_{\ell}^{\mu} &= \beta_L^{b \mu}~, &  \DD_{q\ell}^{s \mu} &= \beta_L^{s \mu}~,
\end{aligned}
\label{eq:U1redef}
\end{align}
where $\beta_{R} \equiv \beta_{R}^{b \tau}$. Note that while~\eqref{eq:U1redef} is just a redefinition of the free parameters
of the effective Lagrangian,~\eqref{eq:U1cond1} give non-trivial constraints. 
We also stress that, beside the overall coupling (encoded in $C_V$) the four combinations of 
couplings in (\ref{eq:U1redef}) indicate the helicity structure of the interactions ($C_S/C_V$) and its
alignment in quark and lepton flavor space ($\lambda^s_q$, $\lambda_\ell^\mu$ and $\DD_{q\ell}^{s \mu}$).

The condition $C_{V_1}=C_{V_3}$, arising naturally in the $U_1$ case, is important to evade the tight constraints on $\cO_{\ell q}^{(3)}$
from $b\to s\nu\bar\nu$ observables and electroweak precision tests~\cite{Buttazzo:2017ixm}. In order to analyze 
charged currents as much independent as possible from other observables, we  take $C_V\equiv C_{V_1}=C_{V_3}$ in the Lagrangian in~\eqref{eq:SMEFTLag}. Similarly, we set $\Gamma_L^{V_1} = \Gamma_L^{V_3} = \Gamma_L$ 
in order to avoid the introduction of redundant parameters as far as  neutral currents are concerned.

\section{The relevant observables}\label{sect:Obs}

\subsection{$b\to c\tau \bar \nu$ rates and polarization asymmetries}

\begin{figure}[t]
\centering
\includegraphics[width=0.475\textwidth]{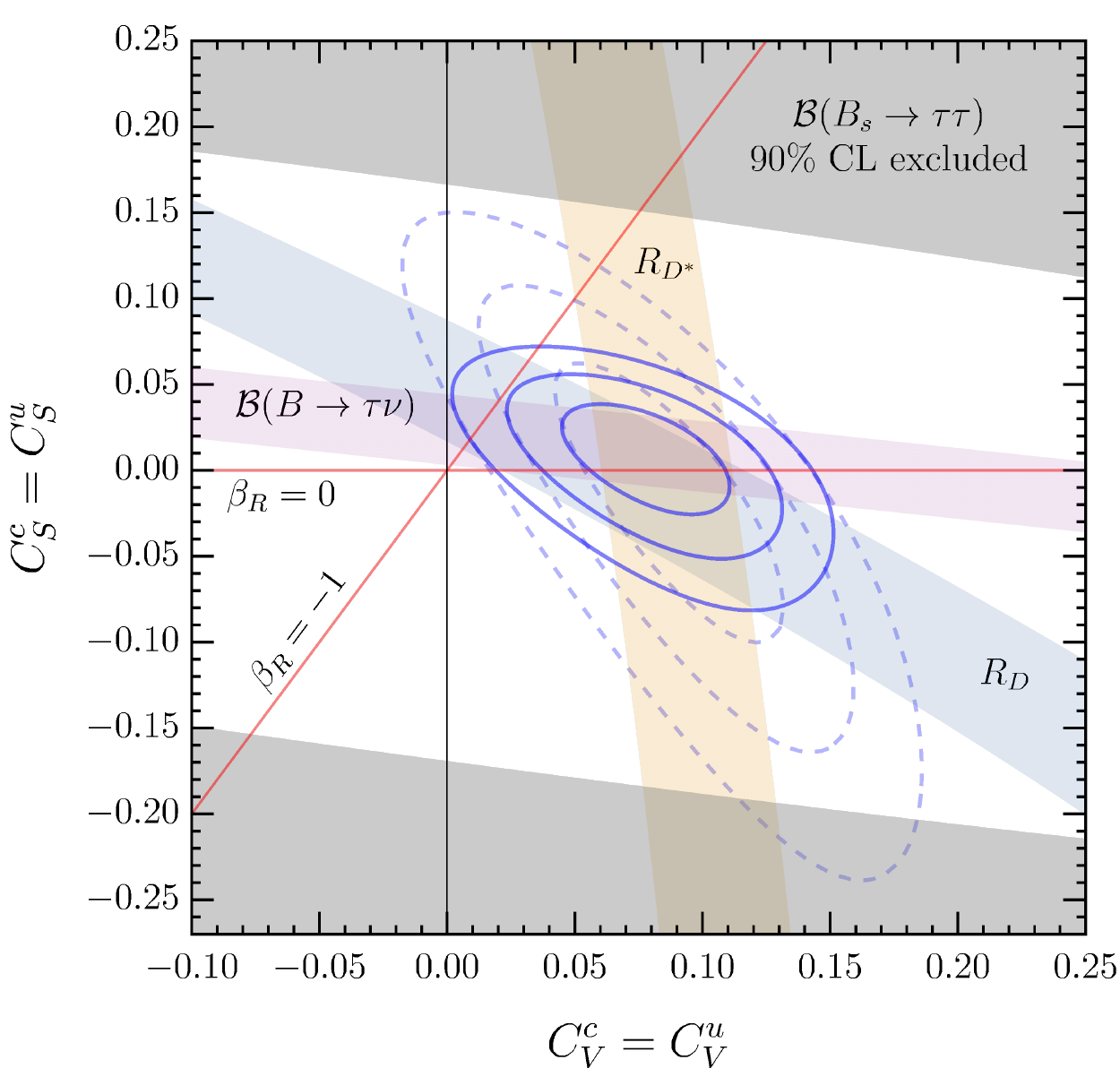}
\caption{\label{fig:CVCS}
Best fit regions in the $(C_S^c,C_V^c)$ plane. The three contours corresponds to $1$, $2$, and $3\sigma$ intervals.
The dashed blues lines take into account only the information from $R_D$ and $R_{D^*}$  (for which we use the HFLAV average~\cite{HFLAV:RDRDs}), whereas the continuous lines 
also include constraints from $b\to u$ observables (see sect.~\ref{sec:b2u} for more details). 
The colored bands correspond to the $1\sigma$ regions defined by each observable. The gray bands show the $90\%$~CL exclusion region from $\mathcal{B}(B_s\to\tau\bar \tau)$ (see sect.~\ref{sec:neutral}). }
\end{figure}

At fixed quark and lepton flavors, the effective Lagrangian in~\eqref{eq:SMEFTLag} depends only on two coefficients.
In the $b\to c\tau \bar\nu$ case, we conveniently re-define them as
\begin{align}\label{eq:CVSc}
\begin{aligned}
C_{V(S)}^{c}  &= C_{V(S)}   \left[ 1+    \lambda_q^{s}  \left(\frac{ V_{cs} }{ V_{cb} } + \frac{V_{cd} }{ V_{cb} } \frac{V_{td}^*}{V_{ts}^*}\right) \right]  \\
 & =  C_{V(S)}  \left( 1-   \lambda_q^s\,\frac{V_{tb}^* }{ V_{ts}^* }  \right)~,
\end{aligned}
\end{align}
where, in the last line, we have used CKM unitarity. When defining $C_{V(S)}^c$,
we have factorized the CKM factor $V_{cb}$, such the that the left-handed 
part of the interactions is modified as
\be
\cA^{\rm SM} \to  (1+C_V^c)  \cA^{\rm SM}~.
\ee
In the absence of the simplifying hypothesis $\Gamma_L^{V_3} = \Gamma_L$, 
one would need to redefine $C_V^c$ replacing $\lambda_q^{s}$ with $\tilde \lambda_q^{s}$.
Employing this hypothesis, as in the leptoquark case, the ratio  $C_{S}^{c}/C^{c}_{V} =C_{S}/C_{V}  $ 
is flavor blind and depends only on the helicity structure of the  NP amplitude.

Using the results in~\cite{Bernlochner:2017jka,Tanaka:2012nw} 
for the $\bar B\to D^{(*)} \ell \bar \nu$ form factors and decay rates, and neglecting the tiny NP corrections in the 
$\ell=\mu,e$ case (see below) leads to the following expression for the LFU ratios, 
$R_H = \Gamma(\bar B\to H \tau \bar \nu)/ \Gamma(\bar B\to H \ell \bar \nu)$, 
\begin{align}
\begin{aligned}
\frac{R_D}{R_D^{\text{SM}}} &\approx |1+C_V^c|^2 + 1.50(1)\,\text{Re}[(1+C_V^c)\,\eta_S\,{C_S^c}^*]  \\
&\quad +1.03(1)\, |\eta_S\,C_S^c|^2\,, \\
\frac{R_{D^*}}{R_{D^*}^{\text{SM}}} &\approx |1+C_V^c|^2 + 0.12(1)\, \text{Re}[(1+C_V^c)\,\eta_S\,{C_S^c}^*]  \\
&\quad + 0.04(1)\, |\eta_S\,C_S^c|^2 \,,
\end{aligned}
\end{align}
where  $\eta_S\approx 1.7$ arises from the running of the scalar operator from the TeV scale down to $m_b$~\cite{Gonzalez-Alonso:2017iyc}.
Updated SM predictions for $R_{D^{(*)}}$ can be found in~\cite{Bordone:2019vic}:
\begin{align}
R_D^{\rm SM}=0.297(3)\,,\qquad R_{D^*}^{\rm SM}=0.250(3)\,.
\end{align}

Current measurements of $R_D$ and $R_{D^*}$~\cite{Aaij:2015yra,Lees:2013uzd,Hirose:2016wfn,Aaij:2017deq,Abdesselam:2019dgh} 
lead to the constraints on  $C^c_S$ and $C^c_V$ shown in Fig.~\ref{fig:CVCS} (dashed contour lines) where, for simplicity, we have 
assumed these couplings to be real. For comparison, the directions corresponding to a pure left-handed 
($\beta_R=0$) or a vector-like interaction ($\beta_R=-1$) for the $U_1$ are also indicated.

Before discussing the impact of additional observables in constraining 
the same set of parameters (under additional assumptions), we stress that once $C^c_S$ and $C^c_V$ have been determined, all the 
other $b\to c\tau \bar \nu$ observables are completely fixed by the $U(2)^5$ invariant structure of $\cL_{\rm EFT}$
and  can be used to test it.  
Particularly interesting in this respect are  the polarization asymmetries, 
\bea
F_L^{D^*} &=& \frac{\Gamma(\bar B \to D_L^* \tau \bar \nu)}{\Gamma(\bar B \to D^* \tau \bar \nu)} \,, \\
P_\tau^{D^{(*)}} &=&  \frac{\Gamma(\bar B \to {D^{(*)}} \tau^{(+)} \bar\nu)-\Gamma(\bar B \to {D^{(*)}} \tau^{(-)} \bar \nu)}{\Gamma(\bar B \to {D^{(*)}} \tau^{(+)} \bar\nu)+\Gamma(\bar B \to {D^{(*)}} \tau^{(-)} \bar\nu)} \,, \qquad \nonumber
\eea
where the $\tau^{(\pm)}$ denotes a $\tau$ with $\pm 1/2$ helicity. We find the following expressions for these observables\footnote{The numerical coefficients 
in (\ref{eq:Pnum}) are compatible with those obtained in~\cite{Iguro:2018vqb}, within the errors, 
and are within $5\%$ of to those obtained in \cite{Blanke:2018yud} (where different form factors have been employed).}
\begin{align}
\begin{aligned} 
\label{eq:Pnum}
\frac{F_L^{D^*}}{F_{L,\text{SM}}^{D^*}}  \approx  \left( \frac{R_{D^*}}{R_{D^*}^{\rm SM}} \right)^{-1} 
 \big(& | 1 + C_V^c|^2   + 0.087(4)\, |\eta_S\,C_S^c|^2   \\
&+ 0.253(8)\, \text{Re}[(1+C_V^c)\,\eta_S\,{C_S^c}^*] \big) \,,\\
\frac{P_\tau^{D}}{P_{\tau,\text{SM}}^{D}}  \approx \left( \frac{R_{D}}{R_{D}^{\rm SM}} \right)^{-1} 
\big(& | 1 + C_V^c|^2    + 3.24(1)\, |\eta_S\,C_S^c|^2   \\
 & +  4.69(2)\, \text{Re}[(1+C_V^c)\,\eta_S\,{C_S^c}^*] \big) \,, \\
\frac{P_\tau^{D^*}}{P_{\tau,\text{SM}}^{D^*}}  \approx  \left( \frac{R_{D^*}}{R_{D^*}^{\rm SM}} \right)^{-1} 
\big(& | 1 + C_V^c|^2  -0.079(5)\, |\eta_S\,C_S^c|^2  \\
  &- 0.23(1) \, \text{Re}[(1+C_V^c)\,\eta_S\,{C_S^c}^*]  \big) \,.
\end{aligned}
\end{align}
Taking $C_{V,S}^c$ real for simplicity, we obtain
\begin{align}
\begin{aligned} 
\label{eq:Pnum}
\frac{F_L^{D^*}}{F_{L,\text{SM}}^{D^*}} & \approx  1 + 0.137(4)\, \eta_S\, C_S^c (1- C_V^c)  + 0.031(1)\, \eta_S^2\, {C_S^c}^2 \,,\\
\frac{P_\tau^{D}}{P_{\tau,\text{SM}}^{D}} & \approx  1 + 3.19(2)\, \eta_S\, {C_S^c} (1-{C_V^c})  -2.59(1) \, \eta_S^2\, {C_S^c}^2 \,, \\
\frac{P_\tau^{D^*}}{P_{\tau,\text{SM}}^{D^*}} & \approx 1 - 0.34(2)\, \eta_S\, C_S^c (1-{C_V^c})  - 0.078(4)\, \eta_S^2\, {C_S^c}^2\,,
\end{aligned}
\end{align}
where~\cite{Bordone:2019vic}
\begin{align}
\begin{aligned} 
F_{L,\text{SM}}^{D^*}&=0.464(10)\,,& P_{\tau,\text{SM}}^{D}&=0.321(3)\,,\\
P_{\tau,\text{SM}}^{D^*}&=-0.496(15)\,.
\end{aligned}
\end{align}
Since the effect of $C_V^c$ is that of rescaling the SM amplitude, all the above ratios are largely insensitive to 
the value of $C_V^c$ and become 1 in the limit $C_S^c \to 0$. 

\begin{figure}[t]
\centering
\includegraphics[width=0.45\textwidth]{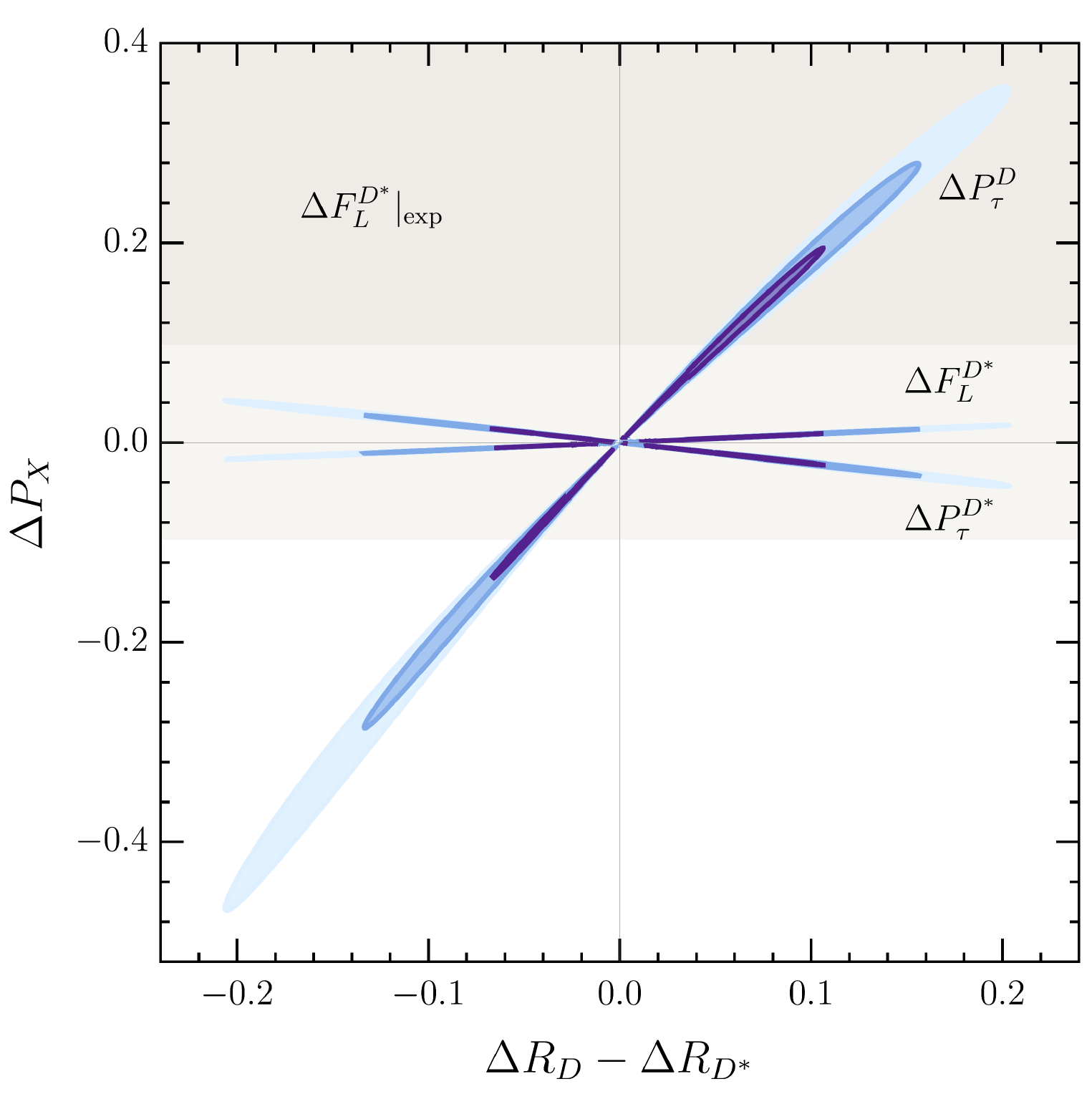}
\caption{\label{fig:Pol}
Deviations of the polarization asymmetries compared to the SM as a function of $\Delta R_D - \Delta R_{D^*}$.
The predictions are obtained  using the fit in Fig.~\ref{fig:CVCS} (continuous lines). In gray, the experimental value of 
$\Delta F_L^{D^*}$ at $1\sigma$ and $2\sigma$. }
\end{figure}

This fact is clearly illustrated in Fig.~\ref{fig:Pol},
where we plot the deviations from unity of the polarization ratios vs.~the difference on the 
two leading LFU ratios (which also vanishes in the limit $C_S^c \to 0$),
\be
\Delta P_X = \frac{ P_X }{P_X^{\text{SM}} } - 1~,  \qquad \Delta R_X = \frac{ R_X }{R_X^{\text{SM}} } - 1~.
\ee
As can be seen, the predicted pattern of deviations is very precise and rather  specific. At present, only $P_\tau^{D^*}$~\cite{Hirose:2016wfn,Hirose:2017dxl} and $F_L^{D^*}$~\cite{Abdesselam:2019wbt} have been measured, still with large uncertainties. As shown in Fig.~\ref{fig:Pol}, it is not possible in our setup to reach the current experimental central value for $\Delta F_L^{D^*}$ (see~\cite{Iguro:2018vqb,Murgui:2019czp} for a similar discussion).

The last  $b\to c\tau \bar \nu$ observable we take into account is 
$\cB(\bar B_c\to \tau \bar \nu )$, that is particularly sensitive to the scalar amplitude.
Despite it will be quite difficult to measure this branching ratio in the future, interesting bounds 
can be derived from the measurement of the $B_c$ lifetime~\cite{Alonso:2016oyd, Celis:2016azn}. 
The expression for this observable reads
\be
\frac{\mathcal{B}(\bar B_c\to \tau \bar \nu)}{\mathcal{B}(\bar B_c\to \tau \bar \nu_{\tau})_{\textrm{SM}}}
 = \left| 1+ C_{V}^c
 + \chi_c\, \eta_S\, C_{S}^c \right|^2~,
\ee
where $\mathcal{B}(\bar B_c\to\tau \bar\nu_{\tau})_{\textrm{SM}}\approx0.02$, $\chi_c = m_{B_c}^2/[m_\tau(m_b + m_c)] \approx 4.3$. 
We find $\mathcal{B}(\bar B_c\to \tau \bar\nu)$ to be at most at the level of $10\%$ for the best fit contours in Fig.~\ref{fig:CVCS}, well below the $\mathcal{B}(\bar B_c\to \tau \bar\nu)\lesssim 30\%$ bound obtained in~\cite{Alonso:2016oyd}.

In principle, additional  probes of the  $b\to c\tau \bar \nu$  amplitude are provided by the 
$\tau/ \mu$ LFU ratios in $\Lambda_b \to \Lambda_c \ell \bar \nu$~\cite{Blanke:2018yud}
 and in $\bar B_c\to \psi \ell \bar \nu$~\cite{Watanabe:2017mip}  
decays. In both cases scalar amplitudes are subleading and, in our framework, one should expect 
an enhancement compared to the SM prediction similar to the one occurring in $R_{D^*}$.  
However, measuring the LFU ratio in  $\Lambda_b \to \Lambda_c \ell \bar \nu$ --where we have a 
precise SM prediction~\cite{Detmold:2015aaa}-- is quite challenging, while in the $\bar B_c\to \psi \ell \bar \nu$ case 
the current SM theory error is well above  the $10\%$ level~\cite{Cohen:2018dgz}.

\begin{figure}[t]
\includegraphics[width=0.42\textwidth]{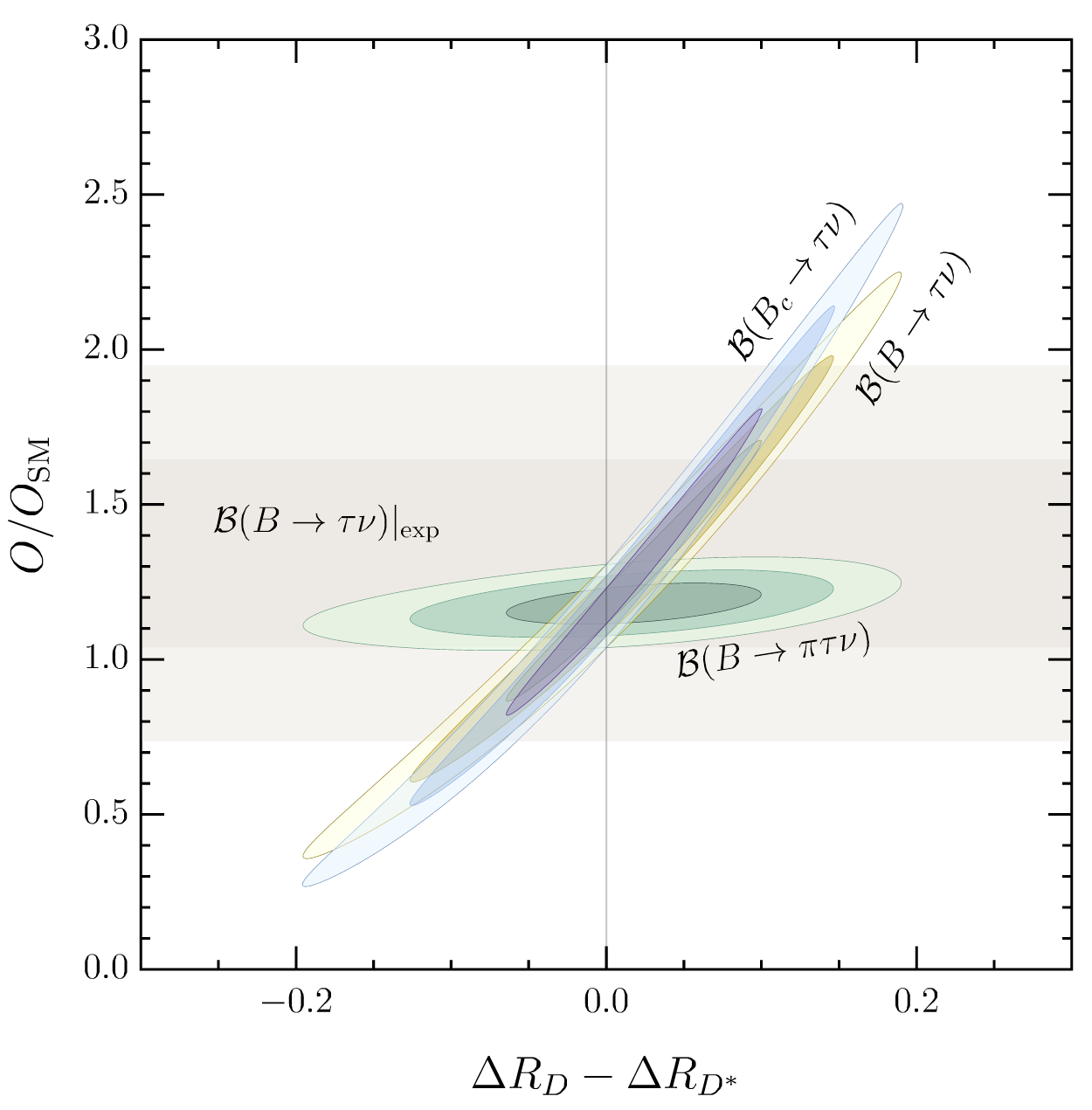}
\caption{\label{fig:BRs}
Predictions for $B(\bar B_c\to \tau\bar\nu)$,  $B(\bar B_u \to \tau\bar\nu)$ and $\cB(\bar B\to \pi \tau \bar \nu)$, all normalized 
to the corresponding SM expectations, as a function of $\Delta R_D - \Delta R_{D^*}$. In gray,
the experimental value of $B(\bar B_u\to \tau \bar\nu)$ at $1\sigma$ and $2\sigma$.}
\end{figure}

\subsection{$b\to u\tau\bar\nu$ transitions}\label{sec:b2u}
The analog of $ C_{V(S)}^{c}$ for $b\to u$ transitions are the effective couplings 
\begin{align}
\begin{aligned}
C_{V(S)}^{u} &=  C_{V(S)}   \left[  1+   \lambda_q^{s} \left(  \frac{V_{us} }{V_{ub}} + \frac{ V_{ud}  }{V_{ub}} \frac{V_{td}^*}{V_{ts}^*} 
  \right) \right]    \\
& = C_{V(S)}^{c}\,,
\end{aligned}
\end{align}
where the result in the second line follows from CKM unitarity. The prediction of same size NP effects, relative to the SM, in $b\to u$ and $b\to c$ transitions is a distinctive feature of the minimally-broken $U(2)^5$ hypothesis.

At present, the most significant constrains on $b\to u\tau\bar\nu$ transitions are derived from $\bar B_u\to \tau \bar \nu$,
whose branching ratio is 
\be
\frac{ \cB(\bar B_u\to \tau \bar \nu) }{ \cB(\bar B_u\to \tau \bar \nu)_{\text{SM}} } = \left| 1 + C_V^u + \chi_u\, \eta_S\, C_{S}^u \right|^2\,,
\ee
where  $\chi_u = m_{B^+}^2/[m_\tau(m_b + m_u)] \approx 3.75$ and $\cB(\bar B_u\to \tau \bar\nu)_{\text{SM}}=0.812(54)$~\cite{Bona:2017cxr}. 
The continuous contour lines in Fig.~\ref{fig:CVCS} 
show the constraints on $C_S^u=C_S^c$ and  $C_V^u=C_V^c$ once the experimental 
data on $R_{D^{(*)}}$ are combined with those on
$\cB(\bar B_u\to \tau \bar\nu)$~\cite{Tanabashi:2018oca} ($1\sigma$ range indicated by the purple band in Fig.~\ref{fig:CVCS}).

In the future, very interesting constraints are expected from $\bar B\to \pi \tau \bar\nu$. Using the hadronic parameters in~\cite{Tanaka:2016ijq,Lattice:2015tia}, we find for 
$R_\pi \equiv \cB(\bar B\to \pi \tau \bar\nu)/ \cB(\bar B\to \pi\ell \bar\nu)$
\begin{align}
\begin{aligned}
\frac{R_\pi}{R_\pi^{\rm SM}} &=  |1+C_V^u|^2 + 1.13(7)\, \text{Re}\left[ (1+C_V^u)\, \eta_S\, {C_S^u}^* \right]   \\
&\quad + 1.36(9) \,|\eta_S\, C_S^u|^2~,
\end{aligned}
\end{align}
where  $R_\pi^{\rm SM} =  0.641(16)$~\cite{Bernlochner:2015mya,Du:2015tda,Tanaka:2016ijq}.
In the limit where quadratic NP effects can be neglected, 
the following approximate relation holds
\be
\frac{R_\pi}{R_\pi^{\rm SM}} ~ \approx ~ 0.75\ \frac{R_D}{R^{\rm SM}_D} + 0.25 \  \frac{R_{D^*}}{R^{\rm SM}_{D^*}} ~,
\ee
which would allow a non-trivial test of the $U(2)^5$ structure of the interactions. 
In Fig.~\ref{fig:BRs} we show the predictions for $\cB(\bar B_u \to \tau \bar\nu)$ , $\cB(\bar B\to \pi \tau \bar\nu)$, and
$B(\bar B_c\to \tau \bar\nu)$, as a function of $\Delta R_D - \Delta R_{D^*}$. 
As shown in the figure, our setup predicts also
\begin{align}
\frac{ \cB(\bar B_u\to \tau \bar \nu) }{ \cB(\bar B_u\to \tau \bar \nu)_{\text{SM}} }\approx \frac{ \cB(\bar B_c\to \tau \bar \nu) }{ \cB(\bar B_c\to \tau \bar \nu)_{\text{SM}} }\,,
\end{align}
where the difference among the two modes arises by sub-leading spectator mass effects 
in the chirality-enhacement factors $\chi_c$ and $\chi_u$.

\subsection{$b\to s  \ell \bar \ell^{(\prime)}$ transitions}\label{sec:neutral}

The $b\to s$  semileptonic transitions  have a rich phenomenology 
and have been extensively discussed in the recent literature. Contrary to the charged-current case, 
here model-dependent assumptions, such as the constraints in \eqref{eq:U1cond1}, 
play a more important role. Rather than presenting 
a comprehensive analysis of the various observables accessible in these modes, our scope here is to focus on:  
i) model-independent predictions related to the minimally-broken $U(2)^5$ hypothesis; 
ii) clean observables controlling the size of the symmetry breaking terms.  

\medskip
\paragraph{\bf $b\to s\tau\bar\tau (\nu\bar\nu)$.}
Under the assumption $C_{V_1}=C_{V_3}$, following from the hypothesis of a $U_1$ UV-completion, NP effects in $b\to s\nu\bar\nu$ transitions are forbidden at tree level. On the other hand, NP contributions to $b\to s\tau\bar\tau$ are almost as large as those in $b\to c\tau\bar\nu$ for $\lambda_q^s=O(0.1)$ (see e.g.~\cite{Buttazzo:2017ixm,Capdevila:2017iqn}). The most relevant observable involving these transitions is $\mathcal{B}(B_s \to \tau\bar\tau)$, which could receive a sizable chiral enhancement:
\begin{align}
\begin{aligned}
\frac{\mathcal{B}(B_s \to \tau\bar\tau)}{\mathcal{B}(B_s \to \tau\bar\tau)_{\rm SM}} &= 
 \left| 1 + \frac{2\pi\,\lambda_q^{s}}{\alpha\, V_{tb}V_{ts}^*\,\mathcal{C}_{10}^{\text{SM}}} \left(C_V + \chi_s\, \eta_S\,  C_S \right) \right|^2  \\
&+\left( 1-\frac{4m_\tau^2}{m_{B_s}^2}\right)  \left|  \frac{2\pi\,\lambda_q^{s} }{\alpha\, V_{tb}V_{ts}^*\mathcal{C}_{10}^{\rm SM}} \,\chi_s\, \eta_S\, C_S  \right|^2\,,
\end{aligned}
\end{align}
where $\chi_s = m_{B_s}^2/[2m_\tau(m_b + m_s)]\approx1.9$ and $\mathcal{C}_{10}^\text{SM} \approx - 4.3$
(see below).
The enhancement of this rate compared to the SM expectation
could reach a factor of $10^2$~($10^3$) for $C_S=0$ ($C_S=2 C_V)$. 
However, the current experimental limit~\cite{Aaij:2017xqt,Bobeth:2013uxa}
\be
\frac{\mathcal{B}(B_s \to \tau\bar\tau)}{\mathcal{B}(B_s \to \tau\bar\tau)_{\rm SM}} < 8.8 \times 10^3\quad (95\%\;\mathrm{CL})~, 
\ee
is still well below the possible maximal enhancement. As a result, at present this observable does not put stringent constraints
on the parameter space of the EFT:  in Fig.~\ref{fig:CVCS} we show the $90\%$~CL exclusion region in the $(C_S^c,C_V^c)$ plane for $\lambda_q^s=3\,|V_{ts}|$. 

As pointed out in~\cite{Crivellin:2018yvo}, a possible large enhancement of the $b\to s\tau\bar\tau$ amplitude can indirectly be tested via the one-loop-induced lepton-universal contributions to $b\to s\ell\bar\ell\;(\ell=e,\mu, \tau)$ in the $\mathcal{O}_9$  direction (see below). This contribution is well compatible and even favored by current data~\cite{Alguero:2019ptt,Aebischer:2019mlg}.

\medskip
\paragraph{$b\to s\mu\bar\mu (e\bar e)$.} FCNC decays to light leptons offer an excellent probe of the $U(2)^5$ breaking terms in the lepton sector. 
These transitions are commonly described in terms of the so-called weak effective Hamiltonian~\cite{Grinstein:1987vj,Buchalla:1995vs}
\begin{align}
\mathcal{H}^{b\to s}_{\rm WET}\supset -\frac{4G_F}{\sqrt{2}}\frac{\alpha}{4\pi}\,V_{tb}V_{ts}^*\sum_{i=9,10,S,P}\mathcal{C}_i^\ell\,\mathcal{O}_i^\ell\,,
\end{align}
with $G_F$ the Fermi constant, $\alpha$ the fine-structure constant and
\begin{align}
\mathcal{O}_9^\ell&=(\bar s \gamma_\mu P_L b)(\bar \ell\gamma^\mu \ell)\,, & \mathcal{O}_{10}^\ell&=(\bar s \gamma_\mu P_L b)(\bar \ell\gamma^\mu \gamma_5\ell)\,,\nonumber\\
\mathcal{O}_S^\ell&=(\bar s P_R b)(\bar \ell \ell)\,, & \mathcal{O}_P^\ell&=(\bar s P_R b)(\bar \ell\gamma_5 \ell)\,.
\end{align}
In the SM, $\mathcal{C}_9^\ell\approx4.1$, $\mathcal{C}_{10}^\ell\approx-4.3$ and $\mathcal{C}_S^\ell=\mathcal{C}_P^\ell=0$. Matching to the Lagrangian in~\eqref{eq:SMEFTLag}, we get ($\mathcal{C}_i=\mathcal{C}_i^{\rm SM}+\Delta \mathcal{C}_i$)
\begin{align}\label{eq:WETmatching}
\begin{aligned}
\Delta \mathcal{C}_9^\mu&=-\Delta \mathcal{C}_{10}^\mu=-\frac{2 \pi}{ \alpha V_{tb} V_{ts}^*} \, C_V\, \DD_{q\ell}^{s \mu}\,  \lambda_{\ell}^{\mu\,*}\,,\\
\mathcal{C}_S^\mu&=-\,\mathcal{C}_P^\mu=\frac{2 \pi}{ \alpha V_{tb} V_{ts}^*} \,\frac{m_\mu}{m_\tau}\, C_S^*\,\DD_{q\ell}^{s \mu}\,s_\tau\,.
\end{aligned}
\end{align}
while the corresponding tree-level effects in the electron sector are negligible. 

One of most relevant observables involving these transitions are the LFU ratios $R_{K^{(*)}}=\Gamma(B\to K^{(*)}\mu\bar\mu)/\Gamma(B\to K^{(*)}e\bar e)$, which are particular interesting due to their robust theoretical predictions: $R_{K^{(*)}}^{\rm SM}=1.00\pm0.01$~\cite{Bordone:2016gaq}. 
In our setup, one gets~\cite{Celis:2017doq,Capdevila:2017bsm}
\begin{align}\label{eq:RK}
R_K&\approx R_{K^*}\approx 1+0.47  \, \Delta \mathcal{C}_9^\mu\,.
\end{align}
The prediction $R_K\approx R_{K^*}$, is a direct consequence of our flavor symmetry assumptions and is independent of the initial 
set of dimension-six SMEFT operators. 
As observed first in~\cite{Hiller:2014ula},
the relation $R_K\approx R_{K^*}$ holds in any NP model where LFU contributions to $b \to s\ell \bar \ell$ decays are induced 
by a left-handed quark current: in our framework this is a direct consequence of the smallness of the flavor-symmetry breaking terms in the 
right-handed sector. From the experimental point of view, this implies that all $\mu/e$ universality ratios in $b \to s$ transitions 
are expected to be the same, provided their SM contribution is dominated by $\mathcal{C}_9$ and/or $\mathcal{C}_{10}$.
In addition to (\ref{eq:RK}), we thus expect\footnote{We 
define the universality ratios as $R(Y_b)_{X_s}=\Gamma(Y_b \to X_s \mu\bar\mu)/\Gamma(Y_b \to X_s e\bar e)$,
assuming a region in $m_{\ell\ell}$ below the charmonium resonances and sufficiently above the 
di-muon threshold $(m_{\ell\ell} \gsim 1$~GeV). }
\be
R_\phi (B_s) \approx 
 R_{\pi K} (B)\approx  R(\Lambda_b)_{\Lambda}\approx  R(\Lambda_b)_{p K}\approx \ldots  \approx  R_K~.
\label{eq:RX}
\ee
Current experimental data hint to  sizable NP effects in $R_K$ and  
$R_{K^*}$~\cite{Aaij:2017vbb,Aaij:2019wad,Wei:2009zv,Abdesselam:2019wac,Aubert:2006vb}, consistent with $R_K\approx R_{K^*}$.

Assuming the NP effect to be the same in $R_K$ and $R_{K^*}$, the combined measurements imply $R_{K^{(*)}}^{\rm exp}=0.80\pm0.05$ (corresponding to  $\Delta\mathcal{C}_9^\mu=-0.43\pm0.11$).
This numerical value provides an important contraint on the size of the leptonic spurion ($\lambda_\ell^\mu$):  since $\DD_{q\ell}^{s \mu} = O( \lambda_q^{s} \lambda_\ell^{\mu})$, 
setting $\lambda_q^{s}=O(10^{-1})$ and $C_V=O(10^{-2})$, as suggested by the $R_{D^{(*)}}$ fit, the value of $R_{K^{(*)}}$ implies $\lambda_\ell^\mu=O(10^{-1})$.
Within the UV leptoquark completion, the fact that $\Delta R_{K^{(*)}}\equiv R_{K^{(*)}}-1  < 0$ allows us also to determine a non-trival relative sign among 
the $U(2)^5$ breking terms:  according to \eqref{eq:U1cond1} one has $C_V>0$, which implies $\DD_{q\ell}^{s \mu}\,  \lambda_{\ell}^{\mu\,*}<0$.

Other data involving $b\to s\mu\bar\mu$ transitions, such as the measurements of 
$P_5^\prime$~\cite{Aaij:2015oid,Wehle:2016yoi,Aaboud:2018krd,CMS:2017ivg} and other differential distributions, 
also deviate from the SM predictions consistently with $R_{K^{(*)}}$, further supporting the hypothesis of $\lambda_q^{s}, \lambda_\ell^{\mu}=O(10^{-1})$. This coincidence in size for quark and lepton spurions points to the interesting possibility of a common origin for the two leading $U(2)^5$ breaking terms.

\begin{figure}[t]
\includegraphics[width=0.42\textwidth]{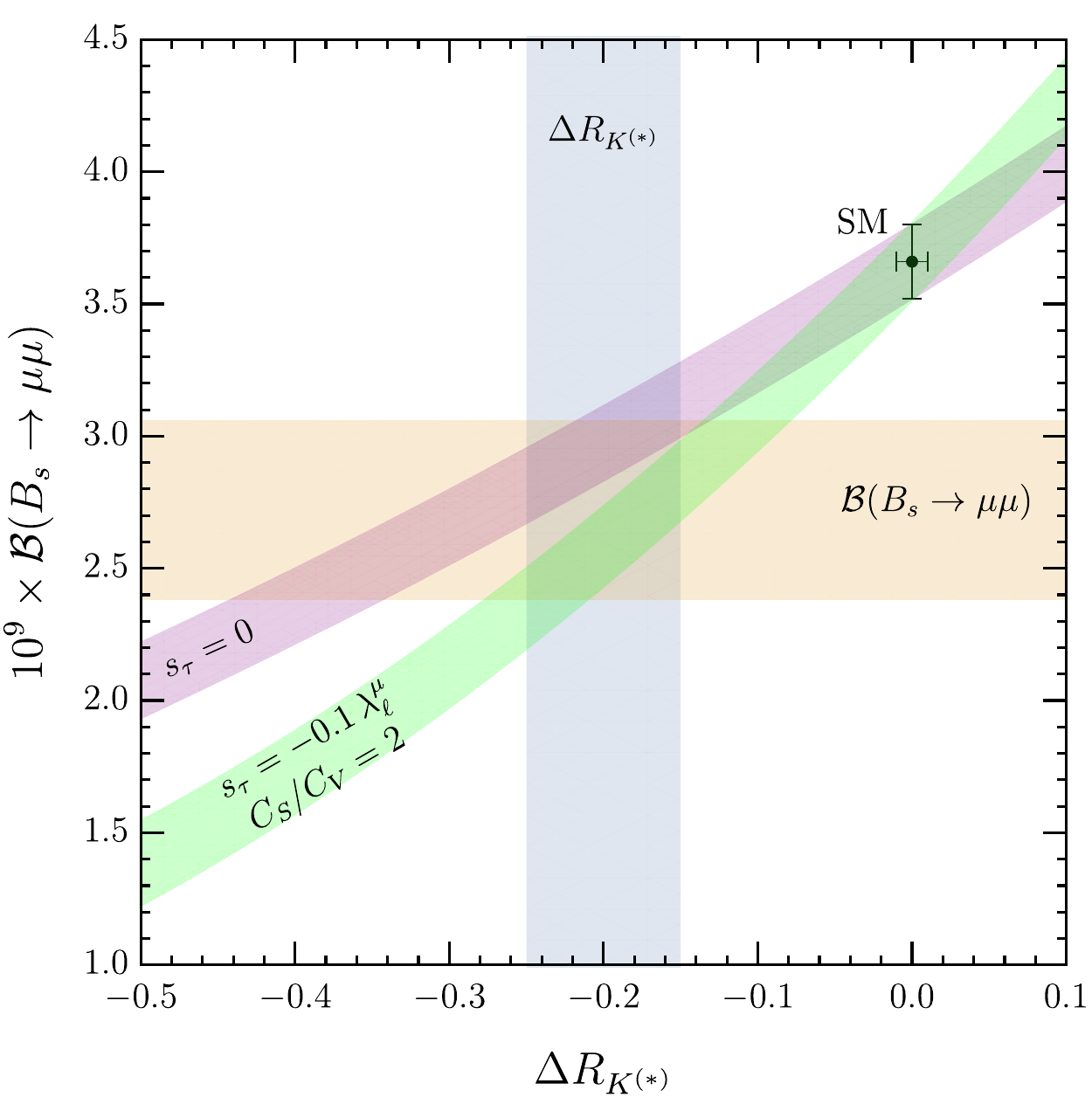}
\caption{\label{fig:Bs2mumu}
Predictions for $\mathcal{B}(B_s\to \mu\bar\mu)$ as a function of $\Delta R_{K^{(*)}}$. The purple and green bands correspond to two different benchmark parameter values. 
The combination of ATLAS, CMS and LHCb measurements of $\mathcal{B}(B_s \to \mu\bar\mu)$, and the combined $R_{K^{(*)}}$ measurement are also shown.}
\end{figure}

\medskip
\paragraph{$B_s\to\mu\bar\mu$.}
Among $b\to s\mu\bar\mu$ transitions, a special role is played by $B_s\to\mu\bar\mu$, where the chiral enhancement of the scalar amplitude
allows us to probe the helicity structure of the NP interaction. The branching ratio normalised to the SM value reads
\begin{align}
\begin{aligned}
& \frac{\mathcal{B}(B_s \to \mu\bar\mu)}{\mathcal{B}(B_s \to \mu\bar\mu)_{\rm SM}} = 
 \left| 1 + \frac{\Delta\mathcal{C}_{10}^\mu}{\mathcal{C}_{10}^{\rm SM}}+\chi_s\,\eta_S\,\frac{m_\tau}{m_\mu}\,\frac{\mathcal{C}_P^\mu}{\mathcal{C}_{10}^{\rm SM}}\right|^2 \\
 &\quad+\bigg( 1-\frac{4m_\mu^2}{m_{B_s}^2}\bigg)  \left| \chi_s\,\eta_S\,\frac{m_\tau}{m_\mu}\,\frac{\mathcal{C}_S^\mu}{\mathcal{C}_{10}^{\rm SM}}  \right|^2\,.
\end{aligned}
\end{align}
Expressing
the deviations in the Wilson coefficients in terms of  $\Delta R_{K^{(*)}}$, 
by means of \eqref{eq:WETmatching} and~\eqref{eq:RK}, leads to 
\begin{align}
&\frac{\mathcal{B}(B_s \to \mu\bar\mu)}{\mathcal{B}(B_s \to \mu\bar\mu)_{\rm SM}} = 
 \left| 1 - \frac{\Delta R_{K^{(*)}}}{0.47\,\mathcal{C}_{10}^{\text{SM}}}\left(1-\chi_s\,\eta_S\,\frac{s_\tau}{\lambda_\ell^\mu}\frac{C_S}{C_V^*}\right)\right|^2 \nonumber \\
 &\quad+\bigg( 1-\frac{4m_\mu^2}{m_{B_s}^2}\bigg)  \left| \frac{\Delta R_{K^{(*)}}}{0.47\,\mathcal{C}_{10}^{\text{SM}}}\,\chi_s\,\eta_S\,\frac{s_\tau}{\lambda_\ell^\mu}\frac{C_S}{C_V^*}\right|^2\,.
\end{align}
Current experimental measurements~\cite{Aaboud:2018mst,Chatrchyan:2013bka,CMS:2014xfa,Aaij:2017vad,CMS:2019qnb} yield 
\be
\mathcal{B}(B_s \to \mu\bar\mu)_{\rm exp} = 2.72(34) \times10^{-9}~,
\ee
which is about $2.6\sigma$ below the SM expectation: 
$\mathcal{B}(B_s \to \mu\bar\mu)_{\rm SM}=3.66(14)\times10^{-9}$~\cite{Beneke:2019slt}.
In Fig.~\ref{fig:Bs2mumu}, we show the predictions for this observable as a function of $\Delta R_{K^{(*)}}$ for $s_\tau=0$ (purple band), and for 
$s_\tau=-0.1\,\lambda_\ell^\mu$ setting $C_S/C_V =2$ (green band). 
As can be seen, the deviations in $R_{K^{(*)}}$ are well compatible with the current experimental values of $\mathcal{B}(B_s \to \mu\bar\mu)$ and, if $C_S/C_V$ is large, small values of $s_\tau$ are favored.

\medskip
\paragraph{$b\to s\tau\bar \mu$.}
As far as  LFV processes are concerned, the most relevant observable is 
\begin{align}\label{eq:Bs2taumu}
\begin{aligned}
\mathcal{B}(B_{s} \to \tau^{-} \mu^{+}) &\approx \frac{\tau_{B_s} m_{B_{s}}  f_{B_{s}}^{2} G_{F}^{2}}{8 \pi}\, m_{\tau}^{2} \left(1 - \frac{m_{\tau}^{2}}{m_{B_{s}}^{2}}\right)^{2} \\
&\quad \times | \DD_{q\ell}^{s \mu} |^2\,\left|  C_V +  2\,\chi_s\, \eta_{S}\,C_S^*\right|^{2} \,. 
\end{aligned}
\end{align}
As in $\mathcal{B}(B_s \to \tau\bar\tau)$, the large chiral enhancement of the scalar contribution make it an excellent probe of the helicity structure of the NP effects. Moreover, this observable provides a direct probe of $\DD_{q\ell}^{s \mu}$. Setting $C_S, C_V=O(10^{-2})$ and $\DD_{q\ell}^{s \mu}=O(10^{-2})$, we find $\mathcal{B}(B_{s} \to \tau^{-} \mu^{+})=\mathrm{few}\times 10^{-6}$, while for $C_S=0$ and the same values of the other NP parameters,  the expected branching fraction is about one order of magnitude smaller. The current experimental limit, 
$\mathcal{B}(B_{s} \to \tau^\pm \mu^\mp)<4.2\times 10^{-5}\;(95\%\;\mathrm{CL})$~\cite{Aaij:2019okb}, is close to the NP predictions when $C_S$ is sizable. Future improvements in this observable will therefore provide very significant constraints.

\subsection{$b\to d \ell\bar\ell $ and other FCNCs} \label{sec:neutralOther}

A key prediction of the  minimally broken $U(2)^5$ framework is that NP effects in $b\to s \ell\bar\ell $ and $b\to d  \ell\bar\ell  $ transitions scale according to the 
corresponding CKM factors. More precisely, defining the effective hamiltonian of the leading $b\to d$ FCNC operators as 
\begin{align}
\mathcal{H}^{\rm b\to d}_{\rm WET}\supset -\frac{4G_F}{\sqrt{2}}\frac{\alpha}{4\pi}\,V_{tb}V_{td}^*\sum_{i=9,10,S,P}\mathcal{\tilde C}_i^\ell\,\mathcal{\tilde O}_i^\ell\,,
\end{align}
where  $\mathcal{\tilde O}_i^\ell =   \mathcal{O}_i^\ell [s\to d]$, then it is easy to check that, because of \eqref{eq:lambda_ratios},
\be
\Delta \mathcal{\tilde C}_{9,10}^\ell  = \Delta \mathcal{C}_{9,10}^\ell~, \qquad  \mathcal{\tilde C}_{S,P}^\ell =  \mathcal{C}_{S,P}^\ell~.
\ee
These relations lead to a series of accurate predictions which could be tested in various $b\to d\ell\bar\ell$ observables.

One of the cleanest test is obtained by means of $B\to\pi\mu\bar\mu (e\bar e) $ decays, where we expect 
\begin{align}
 \frac{\mathcal{B}(B\to\pi\mu\bar\mu)_{[\Delta q^2_{\rm pert}  ]}}{\mathcal{B}(B\to\pi e\bar e )_{[\Delta q^2_{\rm pert}]}}\approx R_{K^{(*)}}\,,
\label{eq:Bpp}
\end{align}
where $\Delta q^2_{\rm pert}$ denotes an interval in $q^2=m_{\ell\bar\ell}^2$ where perturbative contributions are dominant.\footnote{The $q^2$ regions 
where perturbative contributions dominates  over charmonia or light resonance terms 
 are the low-$q^2$ region ($2~{\rm GeV}^2 \lesssim q^2 \lesssim 6~{\rm GeV}^2$) and 
the high-$q^2$ region ($q^2 \gtrsim 15~{\rm GeV}^2$).}
The SM prediction for the rate is $\mathcal{B}(B^+\to\pi^+\mu\bar\mu)_{[1,6]}^{\rm SM}=1.31(25)\times10^{-9}$~\cite{Khodjamirian:2017fxg} 
and $\mathcal{B}(B^+\to\pi^+\mu\bar\mu)_{[15,22]}^{\rm SM}=0.72(7)\times10^{-9}$~\cite{Bailey:2015nbd}, to be compared with the LHCb results $\mathcal{B}(B^+\to\pi^+\mu\bar\mu)_{[1,6]}=0.91(21)\times10^{-9}$ and $\mathcal{B}(B^+\to\pi^+\mu\bar\mu)_{[15,22]}=0.47(11)\times10^{-9}$~\cite{Aaij:2015nea}. These measurements deviate from the SM predictions by $1.2\sigma$ and $2\sigma$, respectively, and are well consistent with~\eqref{eq:Bpp} 
(assuming NP effect in the $e\bar e$ mode to be subleading), although 
they are still affected by large errors. 
Similarly, our framework predicts
\begin{align}
\frac{\mathcal{B}(B_s \to \mu\bar\mu)}{\mathcal{B}(B_s \to \mu\bar\mu)_{\rm SM}}\approx\frac{\mathcal{B}(B_d \to \mu\bar\mu)}{\mathcal{B}(B_d \to \mu\bar\mu)_{\rm SM}}\,.
\label{eq:Bdmm}
\end{align}

Leaving aside $B$ decays, the effective Lagrangian \eqref{eq:SMEFTLag} necessarily imply non-standard effects also in $K$ and $\tau$
semileptonic  decays. Since NP effects in these processes are strongly constrained,  
it is important to check if these constraints limit the parameter space of the EFT. 
As far as $\tau$ decays are concerned, the most stringent constraint is obtained by the non-observation of $\tau\to\mu\gamma$. 
Only the chiral-enhanced contribution to $\tau\to\mu\gamma$, proportional to $C_S$, can be reliably estimated in the EFT, yielding\footnote{The full branching fraction, including both vector and non-chiral-enhanced scalar contributions, was computed in~\cite{Cornella:2019hct}  in a specific $U_1$ UV-completion. There, it was found that the additional contributions are much smaller than the ones quoted here if $C_S\sim C_V$. }
\begin{align}\label{eq:tau2mugamma}
\mathcal{B}(\tau \to \mu \gamma) & \approx  \frac{1}{\Gamma_{\tau}} \frac{\alpha}{256 \pi^{4}} \frac{m_{\tau}^3m_b^2 }{v^{4}}\, | C_S\, \lambda_{\ell}^{\mu\,*}|^2 \,.
\end{align}
Taking $C_S=O(10^{-2})$ and $\lambda_\ell^\mu=O(10^{-1})$, we find $\mathcal{B}(\tau \to \mu \gamma)=\mathrm{few}\times10^{-9}$, which is below current bounds but within the expected Belle~II sensitivity~\cite{Kou:2018nap}. 

Finally, the constraints obtained from $K$ decays do not yield significant bounds to our framework. As in the $b\to s\nu\bar\nu$ case, NP effects in $s\to d\nu\bar\nu$ transitions are forbidden at tree-level if we take $C_{V_1}=C_{V_3}$. On the other hand, $K_L\to \ell\bar\ell^{(\prime)}$ decays receive strong spurion suppressions, resulting in bounds on $C_V$ that are significantly above the preferred values. These are shown in Table~\ref{tab:Kaons}, together with the parametric spurion dependence of the corresponding $d \bar s\to\ell\bar\ell^{(\prime)}$ transition.
For comparison, we stress that the preferred value of $C_V$ emerging from the $R_{D^{(*)}}$ fit in Fig.~\ref{fig:CVCS}, assuming $\lambda_q^s=0.1$, 
is $C_V = C_V^c / ( 1-   \lambda_q^s\,\frac{V_{tb}^* }{ V_{ts}^* } ) \sim (1\div2) \times 10^{-2}$.

\begin{table}[t]
\centering
\renewcommand*{\arraystretch}{1.5}
\begin{tabular}{c|c|c}
Process & Spurion comb. for $d\bar s\to\ell\bar\ell^{(\prime)}$  & Bound on $C_V$\\
\hline\hline
$K^0\to\mu\bar\mu$ & $\Delta_{q\ell}^{s\mu}\,{\Delta_{q\ell}^{d\mu}}^*=\frac{V_{td}}{V_{ts}}\,|\Delta_{q\ell}^{s\mu}|^2$ &  $C_V\lesssim0.3$ \\
$K^0\to\mu \bar e$ & $\Delta_{q\ell}^{se}\,{\Delta_{q\ell}^{d\mu}}^*=s_e\,\Delta_{q\ell}^{s\mu}\,{\Delta_{q\ell}^{d\mu}}^*$ & $C_V\lesssim 0.1\,(\frac{0.2}{s_e})$\\
\end{tabular}
\caption{Constraints from $K$ decays using the analysis in~\cite{Giudice:2014tma}. The bounds on $C_V$ are obtained  setting  $|\Delta_{q\ell}^{s\mu}|=10^{-2}$.}\label{tab:Kaons}
\end{table}

\subsection{Charged-current transitions to light leptons}

The $U(2)^5$ breaking in the lepton sector could in principle be tested also in charged-current decays to light leptons, both in $b\to c$ and $b\to u$ transitions. The most relevant observables in each category are respectively $R_{D^{(*)}}^{\mu e}\equiv\cB(\bar B\to D^{(*)} \mu \bar\nu)/\cB(\bar B\to D^{(*)} e \bar \nu)$ and $\cB(\bar B_u \to \mu\bar\nu)$, whose experimental measurements will be improved at Belle II~\cite{Kou:2018nap}. In contrast to $R_{D^{(*)}}$, scalar contributions to $R_{D^{(*)}}^{\mu e}$ are extremely suppressed due to the $U(2)^5$ flavor symmetry [see~\eqref{eq:betad}]. 

The $\mu/e$  LFU  ratios can be expressed as
\begin{align}
\begin{aligned}
\frac{R_{D^{(*)}}^{\mu e}}{R_{D^{(*)}}^{\mu e\;{\rm SM}}}&\approx |1+ \lambda_\ell^{\mu\,*}\, C_V^{c\mu}|^2 + |\lambda_\ell^{\mu\,*}\, C_V^c|^2 \,,
\end{aligned}
\end{align}
where the first term corresponds to the mode with $\nu=\nu_\mu$ and the second one with $\nu=\nu_\tau$ and where, similarly to $C_V^c$, we have defined
\begin{align}
\begin{aligned}
C_V^{c\mu}&\equiv\lambda_\ell^\mu\, C_V   \left[  1-   \frac{\DD_{q\ell}^{s \mu}}{\lambda_\ell^\mu}\frac{V_{tb}^*}{V_{ts}^*} \right]\,.
\end{aligned}
\end{align}
Taking $\lambda_\ell^\mu=O(10^{-1})$, $C_V^c=O(10^{-1})$ and $C_V^{c\mu}=O(\lambda_\ell^\mu\,C_V^c)=O(10^{-2})$, we find that NP corrections to these observables are at most at the per-mille level, hence beyond the near-future experimental sensitivity. 
This is quite different from what is expected in other NP models addressing the anomalies, such as the scalar leptoquark models considered in~\cite{Dorsner:2017ufx,Becirevic:2018afm}.
 
It is also worth stressing that the phenomenological condition $\DD_{q\ell}^{s \mu}\,  \lambda_{\ell}^{\mu\,*}<0$, required to accommodate $R_{K^{(*)}}$ with $C_V>0$, yields a partial cancellation between the two terms in $C_V^{c\mu}$. As a result of this cancellation, NP effects are typically at the sub per-mille level, hence beyond any realistic sensitivity even in a long-term perspective. Similarly, we find possible NP contributions to $\cB(\bar B_u\to\mu\bar \nu)$ to be at or below the per-mille level, very far from the experimental reach.

\section{Conclusions}
The hints of LFU violations observed in $B$ meson decays have shaken many prejudices 
about physics beyond the SM, opening new directions in model building. 
One of the most intriguing possibilities is the existence of a link between the (non-standard) 
dynamics responsible for these anomalies and that responsible for the fermion mass hierarchies. 
A specific realization of this idea is the hypothesis that, at low energies, the new dynamics 
manifests  via an EFT controlled by an approximate $U(2)^5$ symmetry, 
with leading breaking in specific directions in the $U(2)_q $ and $U(2)_\ell$ subgroups.

In this paper we have explored in generality the consequences of this 
symmetry and symmetry-breaking  hypothesis in (semi)leptonic  $B$ decays,
trying to avoid making additional dynamical assumptions about the origin of the anomalies.
As we have shown, the symmetry hypothesis alone leads to a significant reduction in the 
number of free parameters of the EFT which, in turn, can be translated 
into stringent predictions 
on various low-energy observables. The situation is particularly simple in the case of 
charged currents, where all relevant processes are controlled by two independent combinations 
of effective couplings. The latter can be determined for instance from $R_{D}$ and $R_{D^{*}}$,
leading to a series of unambiguous predictions for 
 $\cB(\bar B_{c,u}\to \ell \bar \nu)$, $\cB(\bar B \to \pi \ell \bar\nu)$, 
 polarization asymmetries in $\bar B\to D^{(*)} \tau \bar \nu$, as well as other processes.
 As shown in Fig.~\ref{fig:CVCS}, the available data on  $\cB(\bar B \to \tau \bar \nu)$
 perfectly support the initial hypothesis. 

In neutral currents, additional combinations of effective couplings appear, 
but also in this case a series of stringent predictions, which are genuine consequences 
of the symmetry and symmetry-breaking  hypothesis alone, can be derived.
The two most notable ones are: i)~the (approximate) universality of the deviations 
from~1 in $\mu/e$  ratios in short-distance dominated $b\to s \ell\ell$ trasitions,  
leading to (\ref{eq:RX}), 
and~ii)~the SM-like CKM scaling of NP effects in $b\to s$ and $b \to d$ transitions, 
which leads to the relations (\ref{eq:Bpp}) and (\ref{eq:Bdmm}).
If the significance of the current anomalies will increase, 
the experimental tests of these relations, which are within the reach 
of current facilities, will provide an invaluable help in clarifying the origin
of this intriguing phenomenon.

\section*{Acknowledgements}

We thank Ryotaro Watanabe for fruitful discussions.
This project has received funding from the  European Research Council (ERC) under the European Union's Horizon 2020 research and innovation programme  under grant agreement 833280 (FLAY), 
and by the Swiss National Science Foundation (SNF) under contract 200021-159720. 
The work of J.F. was also supported in part by the Generalitat Valenciana under contract SEJI/2018/033.
The work of K.Y. was also supported in part by the JSPS KAKENHI 18J01459.

\appendix

\section{Physical parameters in the $U(2)^5$ spurions}\label{app:U2Yukawas}
We describe here how to remove the unphysical parameters in the spurions appearing in~\eqref{eq:U2Yukawas}. Let us focus first on the quark sector. In the exact $U(2)^5$ limit, the Lagrangian presents a
\begin{align}
U(2)_q \times U(2)_u \times U(2)_d \times U(1)_{B_3} 
\end{align}
flavor symmetry. Since, by assumption, this symmetry is only broken by $\DD_{u,d}$ and $\Vq$, field transformations under this symmetry only modify these spurions. In particular, the $U(2)^3$ transformation\footnote{We use the $l$ subscript to denote first- and second-generation fermions.}
\begin{align}
q_l&\to U_Q\, q_l\,, & 
u_l&\to U_U\, u_l\,,  &
d_l&\to U_D\, d_l\,,  
\end{align}
with $U_{Q,U,D}$ being general $2\times2$ unitary matrices, let us bring $\DD_{u,d}$ to the following form
\begin{align}
\begin{aligned}
\DD_u&\to U_Q^\dagger\, \DD_u\, U_U=O_u^\intercal\,\hat\DD_u\,,\\
\DD_d&\to U_Q^\dagger\, \DD_d\, U_D=\hat\DD_d\,.
\end{aligned}
\end{align}
Here $\hat\DD_{u,d}$ are diagonal positive matrices and $O_u$ is an orthogonal matrix. This is the well-known result that in the SM with two families, only the Cabbibo matrix is physical and there are no observable  CP-violating phases. A residual $U(1)_{B_l}$ symmetry, corresponding to baryon number for light quarks, remains unbroken by these spurions. In terms of degrees of freedom,  we started with 8 real parameters and 8 phases in $\DD_{u,d}$, and we used the full freedom of the $U(2)^3/U(1)_{B_l}$ symmetry to remove $11$ parameters. In the end, $5$ real parameters remain: $4$ quark masses and one mixing angle.

The field redefinitions performed so far also redefine the $V_q$ spurion: $V_q\to \tilde V_q\equiv U_Q^\dagger\, V_q$. We can decompose it as
\begin{align}
\tilde V_q&=e^{i\alpha_q}\,|V_q|\,U_q\,\vec n\,, & \vec n&=\begin{pmatrix}0\\ 1\end{pmatrix}\,,
\end{align}
with $U_q$ defined as in~\eqref{eq:Uq}. Without loss of generality, the phase $\alpha_q$ can be absorbed into a redefinition of $x_{t,b}$. Moreover, we can use the $[U_{B_l}\times U(1)_{B_3}]/U(1)_B$ symmetry transformation to remove one phase combination from the two phases in $x_{t,b}$. We use this freedom to set equal phases for $x_b$ and $x_t$. Finally, we can absorb the phase of $y_{t,b}$ into a redefinition of the $t_R$ and $b_R$ fields. This final redefinition  modifies the phase of any $U(2)^3$-preserving NP interaction, but it does not affect the rest of the SM Lagrangian. After all these field transformations, we end up with the following quark Yukawa matrices
\begin{align}
\begin{aligned}
Y_u&=|y_t|
\begin{pmatrix}
\;O_u^\intercal\,\hat\DD_u &&& U_q\,|V_q|\,|x_t|\,e^{i\phi_q}\,\vec n\\
\;0 &&& 1
\end{pmatrix}
\,,\\
Y_d&=|y_b|
\begin{pmatrix}
\;\hat\DD_d &&& U_q\,|V_q|\,|x_b|\,e^{i\phi_q}\,\vec n\\
\;0  &&& 1
\end{pmatrix}
\,,
\end{aligned}
\end{align}
A final $U(2)_q$ transformation, $q_l\to U_q^\dagger\, q_l$, is sufficient to bring these matrices to the form in~\eqref{eq:U2YukawasNonR}. 

The same procedure can be applied to the lepton sector, however in this case there are three differences with respect to the quark case (in the limit of vanishing neutrino masses): i) a subgroup of the $U(2)_\ell\times U(2)_e$ symmetry is enough to make $\DD_e$ diagonal and positive, with no extra orthogonal matrices; ii) the remaining freedom let us reduce $U_\ell$ to an orthogonal matrix, which we denote as $O_e$; iii) the $[U(1)_{L_l}\times U(1)_{L_3}]/U(1)_L$ transformation can be used to remove the phase in $x_\tau$, leaving no free phases. After an appropriate $U(2)_\ell$ transformation, it is straightforward to arrive to the form in~\eqref{eq:U2YukawasNonR}.

\bibliography{references}

\end{document}